\documentclass[
aps,nofootinbib,showpacs,showkeys,preprint
,tightenlines ] {revtex4}

\usepackage{epsf,epsfig,subfigure,
graphicx,amsmath,amssymb}
\usepackage{color}

\newcommand{\dis}[1]{\begin{equation}\begin{split}#1\end{split}\end{equation}}
\newcommand{\be}{\begin{equation}}
\newcommand{\ee}{\end{equation}}

\newcommand{\eq}[1]{Eq.~(\ref{#1})}

\def\bea{\begin{eqnarray}}
\def\eea{\end{eqnarray}}

\begin{document}

\begin{flushright}
{\tt
PNUTP-12-A05
}
\end{flushright}

\title{\Large\bf 130 GeV Fermi gamma-ray line
\\from dark matter decay
}

\author{
Bumseok Kyae$^{(a)}$\footnote{email: bkyae@pusan.ac.kr}
and Jong-Chul Park$^{(b)}$\footnote{email: jcpark@kias.re.kr}
}
\affiliation{$^{(a)}$
Department of Physics, Pusan National University, Busan 609-735, Korea
\\
$^{(b)}$ Korea Institute for Advanced Study, Seoul 130-722, Korea
}


\begin{abstract}

The 130 GeV gamma-ray line based on tentative analyses on the
Fermi-LAT data is hard to be understood with dark matter
annihilation in the conventional framework of the MSSM. We point
out that it can be nicely explained with two body decay of a
scalar dark matter ($\tilde{\phi}_{\rm
DM}\rightarrow\gamma\gamma$) by the dimension 6 operator
suppressed with the mass of the grand unification scale ($\sim
10^{16}$ GeV), ${\cal L}\supset|\tilde{\phi}_{\rm
DM}|^2F_{\mu\nu}F^{\mu\nu}/M_{\rm GUT}^2$, in which the scalar
dark matter $\tilde{\phi}_{\rm DM}$ develops a TeV scale vacuum
expectation value. We propose a viable model explaining
the 130 GeV gamma-ray line.

\end{abstract}


 \maketitle


A thermally produced weakly interacting massive particle (WIMP)
would be the most promising dark matter (DM) candidate explaining
23 percent of the mass-energy density in the present universe
\cite{Jungman:1995df}. It might be closely associated with a new
physics at the electroweak (EW) energy scale beyond the standard
model in particle physics. In this sense, the on-going indirect DM
searches, which cover the EW energy scales, are expected to
provide a hint toward the fundamental theory in particle physics
as well as in cosmology.

Recent tentative analyses~\cite{Bringmann:2012vr, Weniger:2012tx,
Tempel:2012ey} based on the data from the Fermi Large Area
Telescope (Fermi-LAT) \cite{Ackermann:2012qk,Abdo} exhibited a
sharp peak around 130 GeV in the gamma-ray spectrum coming from
near the galactic center (GC). The authors pointed out the
gamma-ray excess could be a result from DM annihilation to a
photon pair.\footnote{In Refs.~\cite{Su:2010qj, Profumo:2012tr},
it is argued that the gamma-ray line can be still explained with
an astrophysical origin, associated with hard photons in the
``Fermi bubble'' regions.} Since DM should carry no
electromagnetic charge,\footnote{The possibility that DM is a
milli-charged particle has been studied in Refs.
\cite{millicharged, 511millicharged}).}
%
the annihilation, $\chi\chi\rightarrow \gamma\gamma$ is possible
only through radiative processes.
If the sharp peak of the gamma-ray around 130 GeV really
originates from DM annihilation, the annihilation cross section
and the mass of DM would be $\langle\sigma
v\rangle_{\chi\chi\rightarrow\gamma\gamma}=(1.27\pm
0.32^{+0.18}_{-0.28})\times 10^{-27}~ {\rm cm^3/s}$ ($2.27\pm
0.57^{+0.32}_{-0.51}\times 10^{-27}~ {\rm cm^3/s}$) and $m_{\rm
DM}=129.8\pm 2.4^{+7}_{-13}$ GeV, respectively, when the Einasto
(NFW) DM profile employed  \cite{Weniger:2012tx}.
It is almost one order of magnitude smaller than the total cross
section for the thermal production of DM needed for explaining the
present DM density ($\sim 10^{-6}~{\rm GeV/cm^{3}}$), which is
about $3\times 10^{-26}~{\rm cm^3/s}$ \cite{Jungman:1995df}.
Indeed, the cross section of order $10^{-27}$ cm/s is much larger
than the expected estimation of one-loop suppressed processes,
assuming a thermal relic DM \cite{Weniger:2012tx}.

On the other hand, at the tree level, DM may annihilate into other
final states, e.g. $W^+W^-, ZZ, b\overline{b}, \tau^+\tau^-,
\mu^+\mu^-$, etc, which can produce secondary continuous
$\gamma$-ray spectrum through hadronizations or final state
radiations. Thus, one can derive the constraints on the DM
annihilation cross sections for those channels from the Fermi-LAT
$\gamma$-ray observation data. Current limits on those
annihilation modes are at the level of $\mathcal{O}(10^{-26}
- 10^{-25})~{\rm cm^3/s}$ for $m_{\rm DM}\approx 130$ GeV.
For more details, see Refs. \cite{GeringerSameth:2011iw,
Ackermann:2011wa, Hooper:2011ti, Cholis:2012am, Cohen:2012me,
Hooper:2012sr}.
Moreover, produced $W$ and $Z$ bosons can also lead to a
sizable primary contribution to the antiproton flux measured by
PAMELA \cite{pamela}, which provides another constraint of
$\mathcal{O}(10^{-25})~{\rm cm^3/s}$ on the DM annihilation into
$W^+W^-$ and $ZZ$ \cite{Belanger:2012ta, Chun:2012yt}.
Consequently, any annihilating DM model to explain the 130 GeV
$\gamma$-ray signal should also satisfy such limits.

Actually, the full one-loop calculations of the neutralino
annihilation into two photons  \cite{Bergstrom:1997fh,
Bern:1997ng, Bergstrom:1997fj} show that the annihilation cross
section of order $10^{-27}~ {\rm cm^3/s}$ is impossible in the
region of 20 GeV -- 4 TeV DM mass in the minimal supersymmetric
standard model (MSSM).
In fact, the neutralino in the MSSM can be annihilated also into
one  photon plus one $Z$ boson through one-loop induced processes,
and this gamma-ray can cause the excess of the observed flux.
Unlike the case of $\chi\chi\rightarrow\gamma\gamma$, the emitted
photon energy is estimated as $E_\gamma = m_{\rm DM}
(1-m_Z^2/4m_{\rm DM}^2)$ for $\chi\chi\rightarrow \gamma Z$. From
the 130 GeV photon energy, hence, the DM mass of 144 GeV is
predicted. [If the $Z$ boson is replaced by an unknown heavier gauge field $X$, the DM mass can be raised more.] Even in such a case, the cross
section of $\chi\chi\rightarrow \gamma Z$ is just about $10^{-28}~
{\rm cm^3/s}$ in the MSSM \cite{Ullio:1997ke,Bergstrom:1997fj},
which is still smaller than $10^{-27}~ {\rm cm^3/s}$.

Renouncing the possibility of thermal production of the neutralino
DM required to explain the observed DM relic density, the
annihilation cross sections $\langle\sigma v
\rangle_{\chi\chi\rightarrow\gamma\gamma}$ and $\langle\sigma v
\rangle_{\chi\chi\rightarrow\gamma Z}$ can be of order $10^{-27}~
{\rm cm^3/s}$ or even larger. In this case, however, the mass
difference between the chargino and the neutralino should be of
order 10 GeV or less \cite{Hisano:2002fk, Belanger:2012ta}.
Moreover, such large neutralino annihilation cross sections for
the one-loop suppressed processes are necessarily accompanied by
much larger annihilation cross section into $W^+W^-$ of order
$10^{-25}~{\rm cm^3/s}$ or even larger \cite{Belanger:2012ta}. As
discussed above, the large annihilation cross section,
$\langle\sigma v \rangle_{\chi\chi\rightarrow W^+W^-}$, is
constrained by the current Fermi-LAT limits on continuum photon
spectrum \cite{Ackermann:2011wa, Hooper:2011ti, Cholis:2012am,
Cohen:2012me, Hooper:2012sr}.

If the neutralino is wino- or higgsino-like, one might think that
the cross section of $\chi\chi\rightarrow\gamma\gamma$ could be
enhanced by nonperturbative effects, called ``Sommerfeld effect.''
It turns out, however, that the cross section $\langle\sigma
v\rangle_{\chi\chi\rightarrow\gamma\gamma}$ cannot reach
$10^{-27}~ {\rm cm^3/s}$, unless the mass of the neutralino is of
TeV or hundreds of GeV scales \cite{hisano03, cirelli07,
Hryczuk:2011vi}, which is exceedingly heavier than 130 GeV.
In addition, the nonperturbative effects on heavy wino- or
higgsino-like DM also enhance the annihilation cross sections into
$W^+W^-$ and $ZZ$, $\langle\sigma v \rangle_{\chi\chi\rightarrow
W^+W^-/ZZ}$ \cite{hisano03, cirelli07, Hryczuk:2011vi,
Chun:2012yt}, which are inevitably constrained by the Fermi-LAT
continuum photon limits \cite{Ackermann:2011wa, Hooper:2011ti,
Cholis:2012am, Cohen:2012me, Hooper:2012sr} and the PAMELA
antiproton flux limits \cite{Belanger:2012ta, Chun:2012yt}.


Thus, the 130 GeV gamma-ray is quite hard to explain with DM
annihilation, if the framework is restricted within the MSSM: we
need to consider a possibility of the presence of a new DM sector,
introducing a new DM and its interactions with ordinary charged
particles.
The basic reason for the difficulty is the charged superparticles'
masses circulating in the loop cannot be light enough to enhance
the cross section, because they should be heavier than the
neutralino DM.
Hence, if a new interaction coupling between a new DM and charged
particles is introduced, which is larger enough than the weak
coupling, we may obtain the required cross section $\langle\sigma
v\rangle_{\chi\chi\rightarrow\gamma\gamma}$ with relatively heavy
(130 + a few$\times$10 GeV) charged  particles in the loop. Of
course, the out-going interaction of the photons should be still
given by the electromagnetic interaction.
In order to reconcile the difference between the demanded cross
sections for 130 GeV gamma-ray by DM annihilation and for the
thermal relic DM, one may introduce two quite different interactions such
that a photon annihilation interaction with $\langle\sigma
v\rangle_{\chi\chi\rightarrow\gamma\gamma} \sim 2 \times 10^{-27}
{\rm cm^3/s}$ is separated from the interaction explaining the
thermal relic with $\sum \langle\sigma v\rangle \sim 3 \times
10^{-26} {\rm cm^3/s}$~\cite{Dudas:2012pb, Cline:2012nw, KYChoi}.
However, we will not pursue such an ambitious job in this Letter.


Instead, in this Letter we will discuss the possibility that the
gamma-ray line at 130 GeV is explained by DM decay. By comparing
the differential photon flux by DM decay ($\Phi_{\rm dec}$) with
that by annihilation ($\Phi_{\rm ann}$)
\cite{Cirelli:2009dv,PalomaresRuiz:2010pn},
\dis{ &~~
\frac{d\Phi_{\rm dec}}{dE_\gamma d\Omega}=\frac{\Gamma}{4\pi}
r_\odot \left(\frac{\rho_\odot}{2m_{\rm DM}}\right)\int_{\rm
l.o.s.}ds\frac{1}{r_\odot}\left(\frac{\rho_{\rm
halo}(r)}{\rho_\odot}\right)\frac{dN_{\rm dec}}{dE_\gamma} ~,
\\
&\frac{d\Phi_{\rm ann}}{dE_\gamma d\Omega}=\frac{\langle\sigma
v\rangle}{8\pi} r_\odot \left(\frac{\rho_\odot}{m_{\rm
DM}}\right)^2\int_{\rm l.o.s.}ds
\frac{1}{r_\odot}\left(\frac{\rho_{\rm
halo}(r)}{\rho_\odot}\right)^2\frac{dN_{\rm ann}}{dE_\gamma} ~, \label{fluxes}}
one can estimate the decay rate $\Gamma$, needed for explaining
the gamma-ray excess, where $dN_{\rm dec(ann)}/dE_\gamma$ is the
differential photon energy spectrum, $\rho_{\rm halo}(r)$ is the
DM halo density profile, $\rho_\odot \approx 0.4$ GeV cm$^{-3}$ is
the local DM halo density, $r_\odot \approx 8.5$ kpc is the
distance from the GC to the Sun and $\int_{\rm l.o.s.}ds$ is the
integral along the line of sight (l.o.s.). The morphology of the
signal from DM decay is linearly proportional to the DM density
profile, while that from DM annihilation has the density square
dependence. Consequently, the decay case tends to show a less
steep increase of the signal towards the GC compared with the
annihilation case, although the morphology of the signal still has
uncertainty by the DM halo density profile itself.
In addition, for decaying DM more $\gamma$-ray flux is generically
expected from the galactic halo compared to annihilating DM.
However, the Fermi collaboration has observed no $\gamma$-ray
excess from the galactic halo: it just reported the lower limits
on the partial DM lifetime $\tau_{\gamma \nu}$
\cite{Ackermann:2012qk}. Although the best-fit values for the
lifetime are in tension with the limit from the Fermi
collaboration, however, the required lifetime to explain the 130
GeV $\gamma$-ray signal marginally satisfies the experimental
limit allowing $2 \sigma$ level error bars
\cite{Buchmuller:2012rc}. Moreover, there still exist the large
uncertainty of the DM distribution around the GC, and also large
statistical and systematic uncertainties at the moment. To confirm
which scenario explains the 130 GeV $\gamma$-ray line, more
improvement in observation is therefore essential in the near
future. In Ref. \cite{Park:2012xq}, it was shown that both of
decaying and annihilating DM explanations similarly give good
$\chi^2$-fits for DM halo profiles more enhanced around the GC
(with $\alpha>1$), compared to the original form of NFW profile
(with $\alpha=1$).


In Eq.~(\ref{fluxes}), we set the DM mass in the decay case as two times
heavier than the annihilation case to obtain the same resulting
gamma-ray fluxes around $E_\gamma = 130$ GeV. Thus, the decay rate
leading to the same gamma-ray flux by the annihilation with
$\langle\sigma v\rangle_{\chi\chi\rightarrow\gamma\gamma}\sim 2
\times 10^{-27}~ {\rm cm^3/s}$ is estimated as
\cite{Ackermann:2012qk,Abdo} \dis{
\Gamma_{\chi\rightarrow\gamma\gamma} \sim 10^{-29}~{\rm s}^{-1}
,
} by which the life time of DM becomes sufficiently longer than
the age of the universe ($\sim 10^{16}$ s). Note that the
required annihilation cross section for $\chi\chi \rightarrow X
\gamma$ is approximately twice of that for $\chi\chi \rightarrow
\gamma\gamma$; the required decay rate for $\chi \rightarrow X
\gamma$ is also approximately twice of that for $\chi \rightarrow
\gamma\gamma$.

If the gamma-ray excess should be explained by DM decay, the sharp
peak of the gamma-ray would imply two body decay of the DM, since
the three body decay would make the spectrum broad and the
intensity much weaker.
In the case of $\chi\rightarrow\gamma X$, the emitted photon
energy is estimated as $E_\gamma = (1-m_X^2/m_{\rm DM}^2) m_{\rm
DM}/2$. For $E_\gamma\approx 130$
GeV, thus, the required DM mass is around $m_{\rm DM} \approx$ 288
(1138) GeV for $X=Z$ (1000 GeV), which is heavier than that in the
annihilation case.
Thus, in the decaying DM case the DM mass
can be much heavier than 260 GeV, say upto  $\mathcal{O}({\rm TeV})$.

In Ref.~\cite{Garny:2010eg}, radiative DM decays to  gamma-ray
were extensively studied. For fermionic DM decay, the author
considered the following renormalizable interactions: \dis{
\label{gaugeInt} &-{\cal L}_{\rm eff}=\overline{\psi}_{\rm
DM}\gamma^\mu\left[g^L_\psi P_L+g^R_\psi
P_R\right]lV_\mu+\overline{N}\gamma^\mu\left[g^L_N P_L+g^R_N
P_R\right]lV_\mu
\\
&\quad +\overline{\psi}_{\rm DM}\left[y^L_\psi P_L+y^R_\psi
P_R\right]l\Sigma+\overline{N}\left[y^L_N P_L+y^R_N
P_R\right]l\Sigma ~+~{\rm h.c.} , } where ``$g$''s and ``$y$''s
denote the coupling constants, and $P_{L,R}$ the projection
operators. $V_\mu$ and $\Sigma$ are superheavy vector and scalar
fields with the masses $m_V$ and $m_\Sigma$, respectively, which
radiatively mediate DM decay. $N$ and $l$ indicate neutral and
charged fermions, respectively. We suppose $m_{\rm DM}<2m_{l}$ to
disallow the three body decays of DM kinematically at tree level.
Note that for producing the photons radiatively, the vector field
$V_\mu$ (and also $\Sigma$) should carry an electromagnetic charge
like the ``$X$'' or ``$Y$'' gauge bosons in the SU(5) grand unified
theory (GUT).
The interactions of \eq{gaugeInt} yield the estimation of the life
time for the fermionic DM $\psi_{\rm DM}$ \cite{Garny:2010eg}:
\dis{ \label{FermionDecay1} \tau_{\psi_{\rm DM}\rightarrow\gamma
N}\approx 1.7\times 10^{27}~{\rm s}~\frac{0.1}{[\sum_l(\eta
g^L_Ng^L_\psi-g^R_Ng^R_\psi)]^2}\left(\frac{260~{\rm
GeV}}{m_{\psi_{\rm DM}}}\right)^5\left(\frac{m_V}{10^{14}~{\rm
GeV}}\right)^4 , } provided $ m_V\ll m_\Sigma$. Thus, the life
time can be of order $10^{29}$ sec. e.g. for a charged gauge boson
slightly heavier than $10^{14}$ GeV. In many GUT models, however,
the mass of the heavy gauge bosons carrying electromagnetic charges should be
well above a few times $10^{15}$ GeV
for longevity of the proton \cite{dim6proton}. Note that in the decaying DM case,
as discussed above, the DM mass can be much heavier than 260 GeV.
In \eq{FermionDecay1}, thus, $m_V$ could be raised upto $10^{15}$
GeV scale for fermionic DM of TeV scale mass. However, it is not
yet enough to reach the conventional SUSY GUT scale ($\approx 2\times 10^{16}$ GeV).
On the other hand, if $m_V\gg m_\Sigma$, the life time of the
fermionic DM becomes \cite{Garny:2010eg} \dis{
\label{FermionDecay2} \tau_{\psi_{\rm DM}\rightarrow\gamma
N}\approx 5.9\times 10^{28}~{\rm
s}~\frac{0.1}{[\sum_l(y^L_Ny^L_\psi-\eta
y^R_Ny^R_\psi)]^2}\left(\frac{260~{\rm GeV}}{m_{\psi_{\rm
DM}}}\right)^5\left(\frac{m_\Sigma}{10^{14}~{\rm GeV}}\right)^4 .
}
Thus, $\Sigma$ with $10^{14}$ GeV mass still affects the gauge coupling
unification, unless $\Sigma$ composes an SU(5) multiplet with
other fields.

One could also explore the possibility of a {\it scalar} DM
decaying to two gammas.  In that case, however, the resulting
effective dimension five operator ($\phi_{\rm DM}
F_{\mu\nu}F^{\mu\nu}/M_*$) should be extremely suppressed  for its
life time of $10^{29}$ s \cite{Garny:2010eg}: the suppression
factor ($1/M_*$) should be much smaller than $1/M_P$, where $M_P$
denotes the reduced Planck mass ($\approx 2.4\times 10^{18}$ GeV).
Thus, let us consider the case that a scalar DM decays to two
photons via the dimension six operator: \dis{ \label{effL} -{\cal
L}_{\rm eff}=c_{\rm eff}\frac{\tilde{\phi}_{\rm
DM}^*\tilde{\phi}_{\rm DM}}{M_*^2}F_{\mu\nu}F^{\mu\nu} , } where
the scalar DM, $\tilde{\phi}_{\rm DM}$ is assumed to develop a
vacuum expectation value (VEV). With \eq{effL}, the decay rate of
$\tilde{\phi}_{\rm DM}$ is estimated as \dis{ \label{d=6decayrate}
&\qquad\qquad\qquad\quad \Gamma_{\tilde{\phi}_{\rm
DM}\rightarrow\gamma\gamma}\approx \frac{c_{\rm eff}^2}{4\pi}
\left( \frac{\langle\tilde{\phi}_{\rm DM}\rangle}{M_*^2} \right)^2
m_{\tilde{\phi}_{\rm DM}}^3
\\
&\approx 0.85\times 10^{-29} ~{\rm s}^{-1}\left(\frac{c_{\rm
eff}\langle\tilde{\phi}_{\rm DM}\rangle}{200~{\rm
GeV}}\right)^2\left(\frac{10^{16}~{\rm
GeV}}{M_*}\right)^4\left(\frac{m_{\tilde{\phi}_{\rm DM}}}{260 ~{\rm
GeV}}\right)^3  \,. }

Now we attempt to achieve this DM decay rate by constructing a simple supersymmetric (SUSY) model.
The scalar DM, $\tilde{\phi}_{\rm DM}$ can be regarded as the scalar component of a superfield $\Phi$.
In this Letter, we simply assume that $\tilde{\phi}_{\rm DM}$ and also its fermionic superpartner $\Phi_{\rm DM}$ are non-thermally produced.
Let us consider the following superpotential:
\dis{ \label{DMsuperPot} W\supset \kappa
N\Phi^2 , }
where $N$ and $\Phi$ are MSSM singlets, and $\kappa$ is a
dimensionless coupling constant. The $R$ and hyper charges of the
relevant superfields are presented in Table \ref{tab:Qnumb}. We
suppose that waterfall fields, which are not explicitly specified
here, decay eventually to the scalar and fermionic components of
$\Phi$ through $N$: $N^c \rightarrow\Phi_{\rm DM}\tilde{\phi}_{\rm
DM}$, $\tilde{N}^* \rightarrow \Phi_{\rm DM}\Phi_{\rm DM}$, and
$\tilde{N}^* \rightarrow \tilde{\phi}_{\rm DM}\tilde{\phi}_{\rm DM}$
by the above $\kappa$ coupling and the corresponding ``$A$-term.''
By the soft SUSY breaking $A$-term corresponding to the $\kappa$
term in \eq{DMsuperPot} and the soft mass terms in the scalar
potential, $\tilde{\phi}_{\rm DM}$ and $\tilde{N}$ can develop
VEVs at the minimum: \dis{ \langle\tilde{\phi}_{\rm DM}\rangle\sim
\langle\tilde{N}\rangle\sim \frac{m_{3/2}}{\kappa}\sim {\cal O}(1)
~ {\rm TeV} .
 }
We assume that the mass of $\tilde{\phi}_{\rm DM}$ determined by
the soft terms is about $260$ GeV. The $A$-term and the VEVs of
$\tilde{\phi}_{\rm DM}$ and $\tilde{N}$ (and also instanton
effects) break U(1)$_R$ completely. However, the above singlet
fields are still hard to be coupled to the ordinary MSSM fields
carrying non-negative integer $R$ charges: since $N$ and $\Phi$
carry {\it positive fractional} $R$ charges, the ordinary $R$
parity violating terms (and also the terms leading to the dimension 5 proton decay), which require the presence of a spurion
field carrying the $R$ charge of $-1$ ($-2$), should be extremely
suppressed in this framework. As a result, such U(1)$_R$ breaking
leaves intact e.g. the proton stability.

%
%

%
%
\begin{table}[!h]
\begin{center}
\begin{tabular}
{c|cccccc} {\rm Superfields} ~\quad & ~\quad $N$ ~\quad & ~\quad $\Phi$ ~\quad & ~\quad $X$ ~\quad &
~\quad $X^c$ ~\quad & ~\quad $Y$ ~\quad & ~\quad $Y^c$ ~\quad
\\
\hline
U(1)$_{R}$  & ~$4/3$ & ~$1/3$ & ~~$1$ & ~~$1$ & ~$2/3$ & ~$4/3$ \quad
\\
U(1)$_{Y}$ [$=$U(1)$_{\rm em}$] & ~$0$ & ~~$0$ & ~$q$
 & $-q$ & $-q$ & ~$q$  \quad
\end{tabular}
\end{center}\caption{$R$ and hyper charges of the superfields.
The ordinary MSSM matter (Higgs) superfields including the
Majorana neutrinos carry the unit (zero) R charges.
}
\label{tab:Qnumb}
\end{table}
%
%

%
\begin{figure}
\begin{center}
\includegraphics[width=0.55\linewidth]{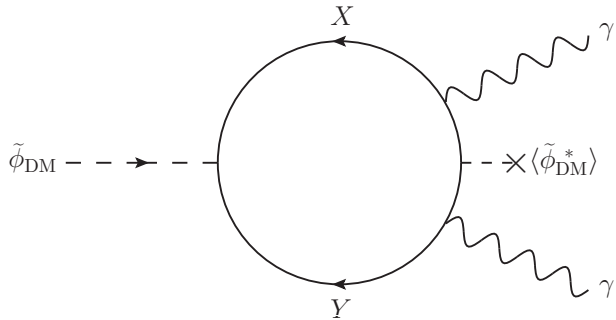}
\end{center}
\caption{
Scalar dark matter decaying to two photons. By radiative
mediations of the superheavy particles $X$ and $Y$, the scalar
dark matter component $\tilde{\phi}_{\rm DM}$ can decay into two
photons with the decay rate of $\mathcal{O}(10^{-29})$ s$^{-1}$. Similar
diagrams contributed by the virtual superheavy scalar partners of
$X$, $Y$ are also possible.} \label{Fig_bino}
\end{figure}



$\tilde{\phi}_{\rm DM}$ cannot be absolutely stable in this
scenario, because of the following additional terms in the
superpotential: \dis{ \label{superXY}
W_{\gamma\gamma}=\lambda_{\gamma\gamma}\Phi XY + M_XXX^c + M_YYY^c
, } where $M_X$ and $M_Y$ are the mass parameters of the GUT scale
($\sim M_{\rm GUT} \approx 10^{16}$ GeV). The U(1)$_{R}$ and
U(1)$_{Y}$ charges of the superfields $X^{(c)}$ and $Y^{(c)}$ are
presented in Table \ref{tab:Qnumb}.
If the model is embedded in an ordinary GUT such as  SU(5) and
SO(10) GUTs, they should carry also SU(3)$_{c}$ and/or SU(2)$_L$
quantum numbers and be accompanied with other (colored) particles
to compose proper irreducible representations and their
conjugations of a GUT. On the other hand, if the gauge group is
given by ``flipped SU(5)'' or just that of the standard model at
the GUT scale, $X^{(c)}$ and $Y^{(c)}$ can still remain as singlet
fields with $q=\pm 1$ without any other supplementary particles
\cite{flippedSU5}.

The DM component, $\tilde{\phi}_{\rm DM}$ eventually decays into
two photons with the desired decay rate. See
Figure~\ref{Fig_bino}. The loop-suppression factor, which gives a
small $c_{\rm eff}$ in \eq{d=6decayrate}, could be easily
compensated by a relatively large VEV of $\tilde{\phi}_{\rm DM}$
($\sim {\cal O}(1)$ TeV), the coupling $\lambda_{\gamma\gamma}$ of
order unity, and a proper choice of charge $q$ such that
\eq{d=6decayrate} can be fulfilled without any serious
fine-tuning. In a similar way, $\tilde{\phi}_{\rm DM}$ can decay
into $\gamma Z$: one photon is replaced with $Z$ in the diagram of
Figure~\ref{Fig_bino}. For $m_{\tilde{\phi}_{\rm DM}} =260$ GeV,
the energy of this single photon is given by
\begin{equation}
E_\gamma = \frac{m_{\tilde{\phi}_{\rm DM}}}{2} \left(
1-\frac{m_X^2}{m_{\tilde{\phi}_{\rm DM}}^2} \right)
 \approx 114 ~{\rm GeV}\,.
\end{equation}
Considering the coupling difference and the reduced phase space
factor, one can easily estimate relative decay rate:
\begin{equation}\label{decayratio}
\frac{\Gamma_{\tilde{\phi}_{\rm DM}\rightarrow\gamma Z}}
{\Gamma_{\tilde{\phi}_{\rm DM}\rightarrow\gamma\gamma}} \approx 0.26\,.
\end{equation}
Since only one photon is produced by the decay mode
$\tilde{\phi}_{\rm DM}\rightarrow\gamma Z$, the final photon flux
is just 0.13 times of the 130 GeV gamma-ray flux. Thus, we predict
another peak around $E_\gamma = 114$ GeV although the 114 GeV
second peak is much less significant than the primary peak around
130 GeV.

On the other hand, the fermionic component of $\Phi$, $\Phi_{\rm
DM}$ cannot decay via a dimension 5 operator: it is because
$\Phi_{\rm DM}$ can decay to a photon plus a neutral fermion
through a dimension 5 operator only when a neutral fermion lighter
than $\Phi_{\rm DM}$ exists as seen in Eqs.~(\ref{FermionDecay1})
and (\ref{FermionDecay2}). However, we do not have such a neutral
fermion. Note that in general a fermionic DM cannot decay to only
two photons via a dimension 6 operator.

Only with the interactions in Eq.~(\ref{superXY}) and the above
field contents, other terms yielding a dimension 6 operator, e.g.
$\Phi^3LH_u$ cannot be generated from the renormalizable
superpotential.
By introducing more fields, however, $\Phi^3LH_u$ might be
induced, since the given symmetries admit it: its  suppression
mass parameter, which is determined by the masses of such
additional fields, would be model-dependent. Even in that case,
the expected signal $\tilde{\phi}^*_{\rm DM}\rightarrow
l^-\tilde{H}^+$ (or $\nu\tilde{H}^0$) is kinematically constrained
by the mass of the Higgsino: if the Higgsino is heavier than
$\tilde{\phi}_{\rm DM}$, $\tilde{\phi}_{\rm DM}$ can decay only to
three or more particles, which is much suppressed if the
suppression factor is of order $M_{\rm GUT}^2$ or heavier. The
other possible decay channel $\tilde{\phi}^*_{\rm DM}\rightarrow
\nu\Phi_{\rm DM}$ might be detectable only in neutrino
observatories such as Super-Kamiokande and IceCube. The current
limits from the Super-Kamiokande data are $\tau_{\rm DM} \approx
10^{24-25}$ sec. \cite{Covi:2009xn}, which is easily evaded if
this decay mode originates from the dimension 6 operator
$\Phi^3LH_u$ suppressed by at least $M^2_{\rm GUT}$. In one year,
the expected sensitivities of IceCube are around $\tau_{\rm DM}
\approx 10^{26}$ sec. \cite{Covi:2009xn, Lee:2012pz}.
Consequently, it is difficult to observe the neutrino signal from
$\tilde{\phi}^*_{\rm DM}\rightarrow \nu\Phi_{\rm DM}$ in near
future. Of course, the bare superpotential can always contain
$\Phi^3LH_u/M_P^2$. However, the $M_P^2$ suppression results in
too weak expected signals.

If  $X$ and $Y$ are accompanied with some other
colored particles to be embedded in certain GUT multiplets,
$\tilde{\phi}_{\rm DM}$ can decay into two gluons, $gg$, through
the similar diagram to that of Figure~\ref{Fig_bino}: two photons
are replaced with two gluons. Then, these gluons can lead to a
sizable primary contribution to the antiproton flux. Thus, this
decay mode $\tilde{\phi}_{\rm DM} \rightarrow gg$ is constrained
by the PAMELA antiproton flux data \cite{pamela}. Even though the
direct constraint on $gg$ channel has not been studied yet and the
antiproton flux constraints depend on the propagation models, we
can estimate the limit as $\Gamma^{-1}_{\tilde{\phi}_{\rm
DM}\rightarrow gg} \approx 10^{26-27}$ s from the limits on the
$W^+W^-, ZZ, hh$, and $q\overline{q}$ channels
\cite{Garny:2012vt}.

In conclusion, we have confirmed that the recently noticed 130 GeV
gamma-ray line based on the Fermi-LAT data is hard to explain with
DM annihilations in the conventional MSSM framework. We raised the
possibility that it originates from two body decay of a scalar
DM ($\tilde{\phi}_{\rm DM}\rightarrow\gamma\gamma$) by a dimension
6 operator suppressed with the mass of the grand unification scale
(${\cal L}\supset|\tilde{\phi}_{\rm
DM}|^2F_{\mu\nu}F^{\mu\nu}/M_{\rm GUT}^2$). The scalar DM needs to
develop a VEV of TeV scale. We proposed a model realizing the
possibility, in which superheavy particles with GUT scale masses
radiatively mediate the DM decay.


\acknowledgments{\noindent  This research is supported by Basic
Science Research Program through the National Research Foundation
of Korea (NRF) funded by the Ministry of Education, Science and
Technology (Grant No. 2010-0009021), and also by Korea Institute
for Advanced Study (KIAS) grant funded by the Korea government
(MEST). }


\end{document}